\documentclass{article}
\usepackage{graphicx}
\usepackage{amsmath}
\usepackage{svrsymbols}
\usepackage{latexsym}
\usepackage{csquotes}
\usepackage{float}
\usepackage{epigraph}
\usepackage{ragged2e}
\usepackage{fontawesome5}
\usepackage{hyperref}
\usepackage{listings}
\usepackage{xcolor}
\lstdefinestyle{mystyle}{
    backgroundcolor=\color{gray!10}, 
    commentstyle=\color{green!50!black},
    keywordstyle=\color{green!80!black},
    numberstyle=\tiny\color{gray},
    stringstyle=\color{orange},
    basicstyle=\ttfamily\footnotesize,
    breakatwhitespace=false,
    breaklines=true,
    captionpos=b,
    keepspaces=true,
    numbers=left,
    numbersep=5pt,
    showspaces=false,
    showstringspaces=false,
    showtabs=false,
    tabsize=2
}
\usepackage{authblk}

\begin{document}

\title{Dark Matter: Explanatory Unification and Historical Continuity}

\author[1,2,3]{Simon Allzén\thanks{\faEnvelope~\href{mailto:simon.allzen@philosophy.su.se}{simon.allzen@philosophy.su.se} or
\faEnvelope~\href{mailto:s.allzen@uva.nl}{s.allzen@uva.nl}}\thanks{The data used for the plots in this paper consists of the bibliographic information from 177,284 papers present in the Astrophysical Data System (ADS) which contain the phrase “dark matter”. See the Appendix for an elaboration of the data processing and analysis.}}
\affil[1]{Department of Philosophy, Stockholm University}
\affil[2]{Institute of Physics, Amsterdam University}
\affil[3]{Vossius Center for the History of Humanities and Sciences, University of Amsterdam}

\date{} 
\maketitle

\begin{abstract}
    \noindent In recent years, the hope to confirm the existence of dark matter by experimentally detecting it has diminished significantly. After more than 30 years of experimental searches, many of the most promising candidates have since been ruled out, leaving the epistemic and scientific condition of dark matter in a state of suspension. In efforts to improve the epistemic justification for the dark-matter hypothesis, physicists have turned to philosophical arguments and historical narratives. In this paper, I explicate two such strategies -- explanatory unification and historical continuity -- applied in the context of dark matter. I argue that greater care and attention should be invested in the explanatory arguments to increase their strength, and that a survey of primary historical sources in astronomy renders the historical evidence for the continuity of dark matter substantially weaker. The quality and rigor of the philosophical and historical arguments which physicists are constructing could be substantially improved by increasing interdisciplinary practices. 
\end{abstract}

\setlength{\epigraphwidth}{.65\textwidth}
\renewcommand{\epigraphsize}{\footnotesize}
\maketitle

\epigraph{\justifying \fontdimen2\font=7pt Is Learning your ambition? There is no royal road;
\noindent \justifying \fontdimen2\font=4.5pt Alike the peer and peasant, Must climb to her abode:
\noindent \justifying \fontdimen2\font=3.5pt Who feels the thirst of knowledge, In Helicon may slake it,\\
\justifying \fontdimen2\font=4pt If he has still the Roman will, \textit{“To find a way, or make it!”}
}{\textit{From: Where there's a will there's a way}\\
\textsc{John Godfrey Saxe, 1884}}


\section{Introduction}
This is a paper on how history and philosophy have been used as partial epistemic justification for the dark-matter hypothesis. Since dark matter has not yet been detected, the paper also engages with the broader question about the principled limits of scientific knowledge.\footnote{Relevant contributions to this intersection of topics include \cite{dawid2013string,dawid2003realism}, \cite{Vanderburgh2014-VANQPE} \cite{kragh2014testability,kragh2017fundamental}, \cite{merritt2021cosmological,merritt2021mond}, \cite{jacquart2021lambdacdm}, \cite{allzen2021scientific}, \cite{SiskaRichard2022mond}, \cite{martens2022integrating}, \cite{antoniou2023robustness}, and \cite{vaynberg2024realism}.} One answer, which also complements the spirit of the above epigraph with the letter, is provided by Carl Sagan. In the essay entitled \emph{Can We Know the Universe?} Sagan reasons about the extent and limits of scientific knowledge:

\begin{quote}
        \small To what extent can we really know the universe around us? Sometimes this question is posed by people who hope the answer will be in the negative [...] And sometimes we hear pronouncements from scientists who confidently state that everything worth knowing will soon be known [...] For myself, I like a universe that includes much that is unknown and, at the same time, much that is knowable. The ideal universe for us is one very much like the universe we inhabit. \cite{sagan1986broca}
\end{quote}
    
\noindent Sagan's perspective expresses an optimistic outlook on the epistemic limits of science: Much of the universe currently unknown to us is nevertheless \textit{knowable}, conditional on our “intellectual zest” (that is, our willingness to fund large-scale scientific projects or the development of advanced technology). The important distinction in this perspective is that the limit is \textit{pragmatic}, not principled. Today, almost 40 years after the publication of Sagan's essay, expressing this optimism borders on naive, primarily because scientists since then have successfully turned many of Sagan's unknowns into knowledge. The success of particle physics, culminating in the Standard Model, is a stand out example of the expansion of scientific knowledge. Ironically, the reasons why particle physics was empirically successful also constitute reasons to doubt that it will continue to be so. The environmental conditions which enabled the success of particle physics were almost ideal, and it stands to reason that the emergence of such an environment is an unusual event, since it depends on the convergence of independent variables -- creative theoretical work, experimental feasibility, and advanced technology -- all reaching maturity simultaneously. The convergence of these variables facilitated a success so comprehensive that theoretical work coupled to achievable empirical boundaries has been largely exhausted, and all particles predicted by that work have been detected. As theorizing has advanced beyond empirically established physics, experimentalists have a hard time following. Theoretical physicist Leonard Susskind describes the situation faced by physicists:

\begin{quote}
    \small The physicist’s guiding star has always been experimental data, but in this respect things are harder than ever. All of us (physicists) are very aware of the fact that experiments designed to probe ever deeper into the structure of matter are becoming far bigger, more difficult, and costlier. The entire world’s economy for one hundred years would not be nearly enough to build an accelerator that could penetrate to the Planck scale. Based on today’s accelerator technology, we would need an accelerator that’s at least the size of the entire galaxy! \cite[261]{Susskind2006tcl}
\end{quote}

\noindent This bleak prediction of the future state of physics is not inevitable, of course, but the reasons to project the trajectory of physics into a future barren of its past empirical success are compelling. This situation also impacts the scientific realism debate, since scientific realists broadly speaking uses empirical success as a truth-marker, meaning that empirically successful theories should be taken to be true, their central terms to refer, and their core entities to exist. The limits of scientific knowledge are under this classical definition directly connected with \textit{empirical success}, which should imply that the future prospects of extending our knowledge through science look slim. 

How is the above relevant in the context of dark matter? The modern understanding is almost unanimously that dark matter is some form of particle, an idea that originated from interdisciplinary scholars in the 1960's and came to fruition in the 1980's. Characterizing dark matter as a particle meant integrating it in the particle physics program and therefore allowing it to be detected. After 30 years of failed detection, in conjunction with the somber outlook for physics described above, the union of dark matter and particle physics has not been good for the former, which inherited the notoriously high evidential standards and expensive machinery from particle physics without getting any of the empirical confirmation.

Attempting to mitigate the impact of possible empirical scarcity in physics, new and systematic \textit{general} frameworks for theory assessment have emerged\footnote{In particular works by \cite{dawid2013string,dawid2015no,dawid2019significance}.} Others use piecemeal rationalist strategies that are \textit{particular} to the situation of the considered theory, attempting to anchor it epistemically or historically in empirically established science. This paper presents examples of the particularist kind, deployed in the case of dark matter. The thematic structure of the paper consists of four parts. Part one illustrates the general demand for alternative modes of justification for theoretical frameworks which extend beyond current empirical boundaries, and ways in which physicists have sought to supply it. Part two outlines the methodological aspects of the empirical project of experimentally detecting a dark-matter particle and explains why the project's lack of success pushed physicists to employ rationalist arguments. Part three concerns the soundness of philosophical arguments used by physicists in support of dark matter, particularly those focusing on justifying dark matter by reference to \textit{explanatory power}. Part four examines the bearing of the assumed historical continuity of dark matter, an assumption which constitutes a central tenet for the scientific pedigree imparted to it by history, as well as for the theoretical continuity and referential stability typically associated with scientific realism about theoretical entities.

\section{Rationalist praxis in physics}

Rationalist strategies in physics involve a range of justificatory constructions, but many of them focus on the properties of the theories themselves. A theory can for example be simple, explanatory, unifying, historically connected, thematically connected, etc. These are properties that a theory can have regardless of the testability of its predictions, making them good candidates for providing justification for theoretical content beyond the empirical horizon. Providing, of course, that these properties can be shown to be of epistemic significance. Later, I will expand on the particular way these properties are manifest in arguments for dark matter, but let us first look at their wider application in physics. This will illustrate that rationalist arguments specific to dark matter are part of a broader pattern of rationalism in physics for purposes of epistemic justification.\footnote{A short disclaimer regarding what follows: the examples of rationalist argument are not chosen because they are \emph{prescriptive} or \emph{desirable}. They are only meant to illustrate the prevalence of rationalist arguments in physics.}

\subsection*{Einstein as a string theorist}
One strategy for justifying a theory is to connect it to the theoretical legacy of a reputable scientist. Greene \cite{greene2000elegant} does precisely this when he highlights the connection between the \textit{aim} of string theory and the \textit{aim} of Einstein's attempt to find a grand unified theory -- unified field theory:

    \begin{quote}
    \small [L]ong after Einstein articulated his quest for a unified field theory but came up empty-handed, physicists believe they have finally found a framework for stitching these insights together into a seamless whole -- a single theory that, in principle, is capable of describing all phenomena. \cite[Preface]{greene2000elegant}
    \end{quote}

\noindent The argument tries to establish a historical link between the aims and aspirations of string theory and Einstein’s pursuit of unification. Such a link positions string theory as the natural successor to Einstein's project, and thereby as continuing on the theoretical trajectory established by perhaps the worlds most revered physicist.\footnote{Or as van Dongen \cite[171]{vandongen2021string} succinctly summarize in his more detailed analysis of this point: “Clearly, Einstein is presented by Greene as both example and justification: the pursuit of string theory is appropriate and honorable, in light of the Einsteinian pedigree of the project”.} Greene’s argument is a species of a broader genus of rationalist strategy to emphasize theoretical continuity between past and present theory. In the case of dark matter, a similar species of this strategy is given in the context of its history.

\subsection*{Unification}
The virtue of unification has a high standing in theoretical physics. To unify disparate phenomena under a single theoretical framework—either by reducing the number of ontological entities or by reducing the number of equations—is taken as a sound guiding principle for physics progression toward truth. Maxwell's equations are often used to exemplify the successful theoretical unification of light, electricity, and magnetism.\footnote{In opposition to this, \cite[132]{maudlin1996unification} argues that Einstein delivered the first true unification because: “in [Special Theory of Relativity] there is truly but one thing: the electromagnetic field tensor”.} Another example of unification as a successful guide is the incorporation of material particles into quantum field theory as quanta of various fields. In the early days of quantum mechanics, electrons and protons were considered to be indivisible particles while photons were considered as manifestations of the quantized electromagnetic field, meaning photons could be destroyed or created. This disparity was by some interpreted as a sign that the theory needed to be unified:

\begin{quote}
    \small It was not long before a way was found out of this distasteful dualism, toward a truly unified view of nature. \cite{weinberg1977search}
\end{quote}

\noindent Weinberg's paper title -- \textit{The Search for Unity: Notes for a History of Quantum Field Theory} -- shows his adoration for unification as a theoretical virtue.\footnote{This adoration was reciprocated two years later when Weinberg, Sheldon Glashow, and Abdus Salam won the Nobel prize in physics for unifying the weak force and the electromagnetic force, since referred to as the electroweak force.} The paper provides a narrative which describes the history of quantum field theory as a \textit{continuous project of unification}, and inserts in this project several prominent scientists of history, including Faraday, Maxwell, Lorenz, and Poincaré. Weinberg's interest in the history of unification in science is motivated by his belief that unification is a kind of yardstick for scientific progress: 

\begin{quote}
     \small The history of science is not merely a tale of intellectual fashions, succeeding one another without direction, but a history of \emph{progress toward truth.} \cite{Weinberg2020Jul}
\end{quote}

\noindent The scope of theory's unification becomes a measurement of its proximity to truth. For Weinberg, unification is a core epistemic concept in science, and he (reluctantly) accepts that this entails defending or explicating the philosophical arguments in favor of it:

\begin{quote}
\small [N]ewton's laws of motion and law of gravity are more fundamental than Kepler's laws of planetary motion. I don't know exactly what I mean by that; presumably it has something to do with the greater generality of Newton's laws [...] we all know what we mean when we say that Newton's laws “explain” Kepler's. We probably could use help from professional philosophers in formulating exactly what that statement means, but I do want to be clear that it is a statement about the way the Universe is, not about the way physicists behave. Weinberg \cite[435]{Weinberg1987Dec}
\end{quote}

\noindent The pursuit of unification is not merely a historical artifact but continues to influence and guide theory construction.\footnote{The close bond between unification and theoretical physics is perhaps best captured by the desire to have: “all of physics reduced to a formula so elegant and simple that it will fit easily on the front of a T-shirt” \cite[21]{lederman2006god}.} Its epistemic value is, however, an addition to the epistemic value of the phenomena it unifies; as the difference between the epistemic value of its empirical parts and their unified sum. As such, unification can bolster theoretical plausibility in contexts that lack access to conclusive empirical confirmation. Later, we will see how this becomes particularly relevant in the context of dark matter, the existence of which is frequently justified by reference to its ability to unify unexplained astronomical phenomena under a single framework.

\subsection*{Simplicity}
\noindent The last example of rationalist arguments in physics is provided on behalf of Einstein himself. In his quest to construct a theoretical framework unifying general relativity with electromagnetism, Einstein placed little importance on empirical evidence, a sentiment he articulated in a 1954 letter to David Bohm:

\begin{quotation}
    \noindent \small I believe that these laws [of unified field theory] are logically simple and that the faith in this logical simplicity is our best guide, in the sense that it suffices to start from relatively little empirical knowledge. \textit{Einstein, 1954}. From \cite{van2010einstein} [181]
\end{quotation}

 \noindent Einstein emphasizes the virtue of logical simplicity for the task of guiding the theoretical progression towards \textit{unification}. Hence, he provides an example of the close connection between the ideas of simplicity, unification, and truth. For him, simplicity is manifested as syntactical economy, but in general simplicity may also refer to ontological parsimony.\footnote{Sometimes, the distinction is merely semantic. See \cite{Vanderburgh2014-VANQPE} for an excellent exposition on the use of simplicity arguments in the dark matter context.} 

In the above examples, historical analysis and philosophical arguments have been used by physicists to assess a theory without referencing any empirical evidence within the theory's own domain. For philosophers and historians of physics, the interest to engage in developing strategies of non-empirical theory assessment is both welcome and worrisome. Welcome, because collaborating on these issues is more likely to generate qualitative results. Worrisome because physicists, rather than dividing the labor among the available expertise, have taken it upon themselves to be historians and philosophers.

\section{Empiricism and the detection of dark matter}
\begin{figure}[ht]
    \centering
    \includegraphics[width=0.8\textwidth]{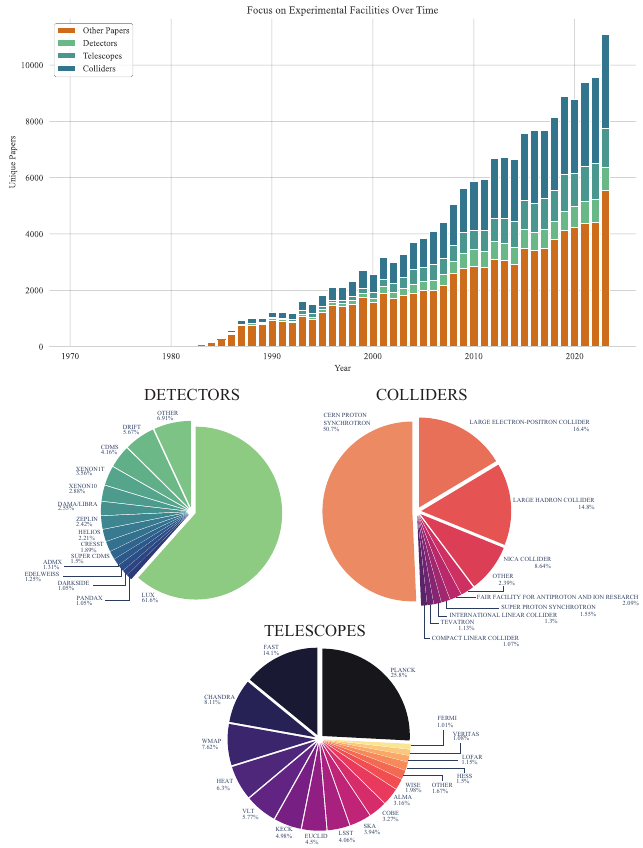}
    \caption{\textbf{Top:} All papers 1970 - 2023, proportion of papers referencing experimental facilities highlighted in stack. \textbf{Bottom:} pie charts of composition of experimental facilities}
    \label{experiments}
\end{figure}

Before examining the rationalist arguments for the reality of dark matter, let us briefly consider the situation that motivated physicists to employ them. Before 1980, dark matter was a peripheral theory at best. Through a series of developments from the mid 1970's through the 1980's, \textit{astronomical} dark matter -- taken to be low-luminous matter effecting galaxy dynamics -- became firmly entrenched as a scientific hypothesis, and the question of its \textit{nature} had by the late 1980's become a matter for particle physicists. For most physicists, dark matter was a particle and because particles can be detected, astronomers and cosmologists were joining the epistemic practices of particle physics. Achieving the revered $5\sigma$ detection would not only entail empirically confirming dark matter, but would also elevate the scientific stature of astronomy and cosmology.\footnote{The ‘age of precision cosmology’ had not yet severed cosmology from its philosophical roots. See \cite{DeSwart2020} for the evolution of cosmology in the mid-19th century.} Detecting dark matter has been optimistically pursued ever since. Figures \ref{experiments} and \ref{HEP} show the various ways in which particle physics has influenced dark matter research. 

The optimism driving these large-scale scientific experiments was, in addition to believing that dark matter is a particle, rooted in the success of particle physics. Its incredible track record in detecting particles reinforced the feasibility of discovering dark matter by detection. Here, cosmologist Carlos Frenk expresses his optimism in an interview:

\begin{quote}

    \small Frenk is willing to bet that scientists will identify dark matter within five years. “Technology has improved and now there is genuine expectation that discovery may be just around the corner,” he said in a telephone interview. “There's a feeling in the air.” \cite{JohnMangels2009Mar}
\end{quote}

\noindent A decade later, astrophysicist Dan Hooper reflected in Time Magazine about his own credences regarding the prospect of detecting a dark-matter particle at the time:

\begin{quote}
    \small If the dark matter is indeed made up of WIMPs, then it should be possible to conduct experiments that could directly detect and measure individual particles of this substance. [...] In fact, I made a bet in 2005 that dark matter particles would be discovered within a decade. I lost that bet. \cite{Hooper2019Dec}
\end{quote}

\noindent The motivation is clear from the initial conditional statement. If dark matter were a WIMP, there were compelling theoretical reasons to expect its detection to be a matter of \textit{when}, not \textit{if}. Figure \ref{experiments} highlights the increasing focus on experimental and observational detection, primarily experimental. We can see how almost half of all research papers written since 2010 refers to detectors and colliders, lending support to the ubiquity of viewing dark matter as a particle, as well as to the hope of detecting it. Although some experiments have advanced detection technology and occasionally produced signals initially interpreted as potential dark matter, later revealed to be background noise, none have succeeded in achieving a direct empirical detection. Nevertheless, these efforts have significantly narrowed the range of viable theoretical candidates, effectively constraining theory space. In this way, dark matter detection programs have complemented the earlier cosmological and astronomical reasoning by refining the parameter space for potential dark matter models. Still, given the ambition to secure a $5\sigma$ detection to secure empirical confirmation, it is difficult to regard this incremental progress as definitive evidence. Here is Hooper again commenting on the precarious situation:

\begin{figure}[ht]
    \centering
    \includegraphics[width=0.8\textwidth]{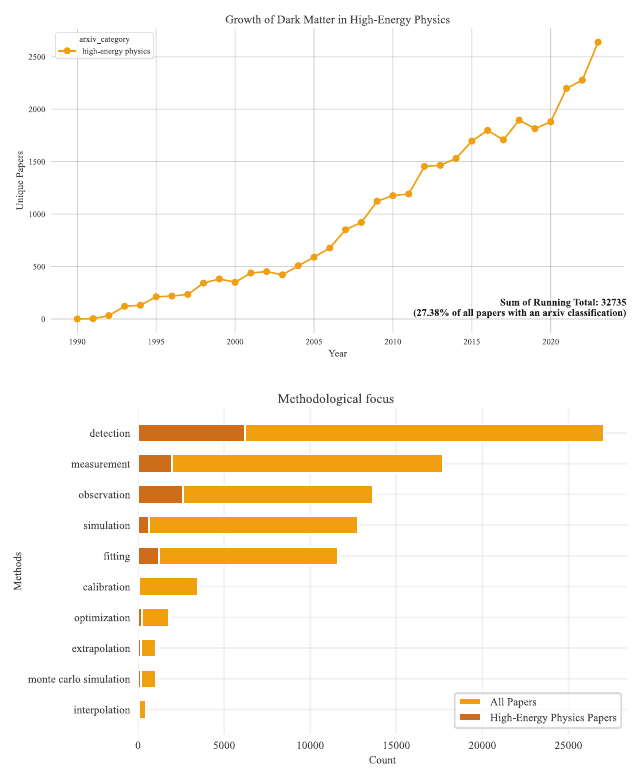}
    \caption{\textbf{Top:} Number of dark matter papers primarily categorized as \textit{high-energy physics}. \textbf{Bottom:} Paper count of methodological terminology, showing the strong presence of detection in the literature}
    \label{HEP}
\end{figure}

\begin{quote}
\small Our failure to detect particles of dark matter has had a palpable effect on the scientific community. Although it remains the case that a discovery could still plausibly lie right around the corner, most of us studying dark matter today will acknowledge that many of our favorite dark matter candidates should have been detected by now. \cite{Hooper2019Dec}
\end{quote}

\noindent Bertone and Tait share this concern, emphasizing that the absence of evidence is not due to lack of trying, given the immense effort dedicated to the search for dark matter:

\begin{quote}
    \small There is a growing sense of ‘crisis’ in the dark-matter particle community, which arises from the absence of evidence for the most popular candidates for dark-matter particles such as weakly interacting massive particles, axions and sterile neutrinos despite the enormous effort that has gone into searching for these particles. \cite{2018Bertone_Tait}
\end{quote}

\begin{figure}[ht]
    \centering
    \includegraphics[width=0.9\textwidth]{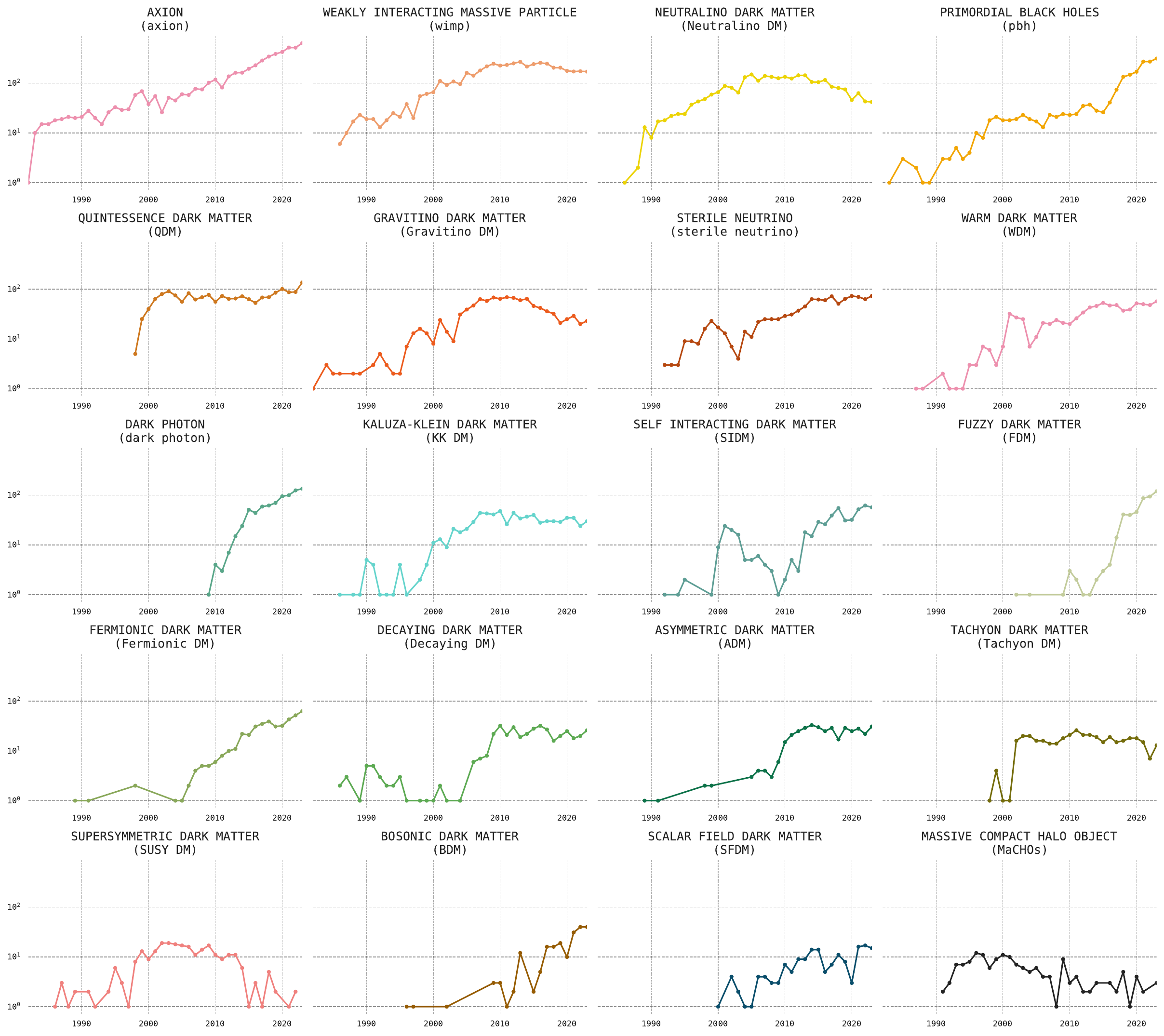}
    \caption{Plot showing the shifting landscape of theoretical models over time, e.g., the rise and fall of WIMPs’, the emergence of axions and primordial black holes}
    \label{dm_models}
\end{figure}

\noindent From a purely quantitative perspective, Figure \ref{dm_models} supports Hooper's assessment of WIMPs as dark-matter candidates. A peak in published papers referring to WIMPs can be observed around 2010, after which interest appears to split: on the one hand, toward increasingly low-mass candidates such as axions (in the lower-mass range starting at $\approx 10^{-50} \text{eV}$), and on the other, toward higher-mass candidates like primordial black holes. The remarkable success of detection methods in particle physics provided the necessary impetus to focus collective efforts on dark matter detectors. Alas, this success has not extended to the detection of dark matter itself. The initial optimism that particle physics could confirm the existence and nature of dark matter now seems to have run its course. This brings us to the core of the paper: the philosophical arguments and historical narratives that have been used to bolster the epistemic justification for dark matter.

\section{Explanation and dark matter}
One of the primary philosophical arguments given in support of the dark matter hypothesis centers on its explanatory powers. Philosophically, such arguments take the explanatory powers of a theory as a \textit{truth conducive} property of the theory. Sometimes, it can be hard to precisely identify the kind of explanatory argument employed, as explanatory reasoning may be used in conjunction with other modes of inference. In the following, a case of explanatory reasoning is interwoven with an argument from analogy.

\begin{quotation}
    \small There, in principle at least, could be an explanation for all of our observations that does not involve the existence of atoms, or planets, or stars. In this sense, any scientific conclusion is provisional. But the fact is that our atomic theory of matter does a \textit{very} good job of explaining a huge variety of data. And so does the existence of cold dark matter.\footnote{Dan Hooper, personal correspondence.}
\end{quotation}

\noindent The analogy aims to demonstrate a parallel between belief in atoms and belief in dark matter with respect to their explanatory power, specifically their ability to provide a single unifying explanation for what would otherwise be disjoint phenomena. The hidden conditional can be explicated along the following lines: if we believe in atoms partly because they, if real, explain a variety of data, then we should believe in the reality of dark matter partly because it, if real, explains a variety of data. Taken as an analogy, the argument is unsuccessful. The belief in atoms arose primarily from experimental confirmation, such as repeated variations of measurements of Avogadro's number (a value that only made sense if atoms were real), and Perrin's experiments to test Einstein's theory that Brownian motion was caused by atoms. Although it is true that the reality of atoms also \emph{explains} many phenomena, the theory of atomism was ultimately accepted because it made \emph{predictions} that were experimentally tested and confirmed. The analogy with dark matter breaks down because, despite its explanatory scope, dark matter has not yet provided comparable predictive success or experimental confirmation.

To succeed, the argument must contain an additional premise: that theories which are explanatorily powerful are more likely to be empirically confirmed. The argument would then be that we ought to believe in dark matter because of its explanatory power, and we know from history that theories with high explanatory power are more likely to become confirmed, as evidenced by the theory of atomism. The structure of such an argument would provide the necessary link between the desiderata, empirical confirmation, and a theoretical property, explanatory power. It then follows that we could (should) increase our credence in the existence of dark matter, even in the absence of de facto empirical confirmation. Arguments of this sort are the bread and butter of philosophy of science, so Hooper is effectively doing philosophy here. I want to caution the reader not to draw the conclusion that these philosophical arguments are decoupled from the empirical domain. While it is true that a theory's explanatory power is inert in an experimental setting, it is nevertheless directly coupled to the empirical. The ability of a theory to explain, unify, or accommodate phenomena naturally connects it to empiricism, since empirical phenomena are the very things being explained. The non-empirical aspect of explanation arises because it is a generic property of the theory itself. There is no empirical basis for discriminating between two explanations of the same phenomena -- if two theories with mutually exclusive ontology offer adequate explanations for some phenomena, there is no \textit{empirical} test to determine which explanation is the 'real' one. Another example of explanatory reasoning, mixed with history and analogy, comes from the following:

\begin{quotation}
    \noindent \small Dark matter has a venerable history. One could even cite Solar System arguments for dark matter, including anomalies in the orbit of Uranus and the advance of Mercury’s perihelion. One led to the discovery of a previously dark planet, Neptune, the other to a new theory of gravitation. \cite[Preface]{BertoneSilk}
\end{quotation}

\noindent The context for the above quote is the discovery of Neptune, which was inferred to be the best explanation of the irregular orbit of Uranus given the fixed background knowledge of Newtonian gravity. Since Neptune was inferred based on the dynamic effects it had on Uranus, and because Neptune was not directly visible, the authors mean that it can be taken as an analogous case of dark matter. It is common in philosophy of science to take the discovery of Neptune as support for the reliability of explanatory inferences. I must confess that I cannot comprehend why the authors included Mercury's perihelion. The “dark matter” inferred to explain the perihelion of Mercury was, as in the Neptune case, taken to be a planet -- Vulcan. But Vulcan was famously proven to \textit{not} exist. In fact, the case of Mercury's perihelion is routinely used to argue \textit{against} the reliability of explanatory inferences. If anything, the case of Mercury's perihelion supports abandoning current theory (Newtonian gravity) in favor of a new theory (general relativity). If we follow the logic of \cite{BertoneSilk}, this means that we should \textit{not} add entities to explain anomalous dynamics, but instead look for new theoretical models. Effectively, it becomes an argument for Modified Newtonian Dynamics (MOND), which is currently the only theoretical rival to dark matter.

\cite{einasto2010dark} -- partially responsible for recognizing the significance of dark matter cosmologically -- refers to the simplicity and elegance of the dark matter hypothesis, and likens the methodological enterprise of confirming dark matter to the “deductions” of Sherlock Holmes:

\begin{quotation}
    \noindent \small [T]he DM paradigm is remarkably simple: one just needs an additional cold collisionless component that interacts only through gravity. Once this component is accepted, a host of apparent problems, starting from galaxy and galaxy cluster scales and extending to the largest scales as probed by the large scale structure and CMB, get miraculously solved. So in that respect one might say that there is certainly some degree of elegance in the DM picture. [...] The search of dark matter can also be illustrated with the words of Sherlock Holmes “When you have eliminated the impossible, whatever remains, however improbable, must be the truth” \cite[14]{einasto2010dark}
\end{quotation}

\noindent Einasto alludes to simplicity and theoretical elegance as the explanatory virtues by which we ought to accept dark matter. He also refers to \textit{eliminative reasoning} as grounds for belief. The latter is dubious, since it depends on which level of discrimination one should apply this reasoning. At the level of \emph{paradigm}, eliminating the impossible would require proof that theories seeking to explain dark phenomena by modifying gravity are impossible.\footnote{“Dark phenomena” is taken from \cite{MARTENS2020237} and denote the observed phenomena not compatible with the predictions generated by GR assuming that only baryonic matter exists.} This has, to my knowledge, not been done, and in addition it is unclear what it would even mean for a theoretical paradigm to be deemed impossible. In opposition to this, \cite[12]{Turner_2022} “MOND posits that $F\simeq ma^2/(a + a_0)$, so that for $a \ll a_0, F \alpha a^2$, but can’t account for cluster dark matter and makes no other predictions. In short, it can’t be falsified”. At the intra-paradigm level, I assume that Einasto is referring to the practice of eliminating dark-matter candidates until only one remains. This strategy is unsound for at least two reasons. First, theoretical candidates that \emph{do} make testable predictions are more vulnerable to elimination, since they are the only candidates that can be empirically ruled out. It is perfectly plausible that only candidates \textit{without} testable predictions will survive the elimination process, so if elevated to principle, we are stacking the deck in favor of candidates which make no predictions within a reasonable empirical boundary. To reward a theoretical model \textit{because} it makes no empirically accessible predictions appears almost like inverse falsification. Second, what reasons do we have to believe that a single unique candidate remains once the process of elimination is complete? In short, the part 'whatever remains' could end up empty. \footnote{This resembles the problem of unconceived alternatives, which stresses the fact that theory space may contain a huge number of viable theories which has not been considered. The point being that the extent of this theory space is difficult to assess.}

In addition, while a professional physicist might better evaluate Einasto’s claim that the dark matter paradigm is simple, it is worth noting the following: the detection and classification of the constituents of \textit{visible} matter required an extraordinary amount of time and effort, culminating in the Standard Model of elementary particles. That model is built on the highly intricate interactions and dynamics of quantum field theory. However, even with its complexity, it accounts for only 20\% of the mass of the universe. Against this backdrop, the assumption that a model capable of explaining the remaining 80\% of \textit{invisible} matter should be simple seems, at best, overly optimistic and, at worst, a poor assessment.

\subsection{The limits and potential of explanatory arguments}

Philosophers of science mostly agree that the bare fact that a theory \textit{can} explain a huge variety of data does not in and of itself entail its truth. Here, just as in Hooper's case, an additional component is needed. Philosophers know the inferential step from a theory's explanatory power to its truth as \textit{inference to the best explanation}, or IBE. In short, explanatory arguments tells us that if we have a set of unexplained (independent) phenomena that can be explained by the introduction of a hypothesis, we ought to increase our belief in that hypothesis. Much like induction, IBE is an \textit{ampliative} inference, meaning that its conclusions necessarily logically extends beyond the information contained in its premises.\footnote{See \cite{laudan1981confutation,Van1989las,fine1991piecemeal,lyons2006scientific} for criticism of IBE. See \cite{psillos1999scientific,Psillos2009}, \cite{douven2002testing}, \cite{vickers2019towards}, and \cite{allzen2022unobservable} for arguments in support of IBE.} Naturally, conclusions from ampliative arguments come without guarantees of truth, but explanatory inferences may still be sufficiently reliable to be epistemically significant. If so, they \textit{can} be used to justify the existence of dark matter, at least to a certain degree: the dark matter hypothesis \textit{is} the best explanation of what would otherwise be unrelated phenomena, and so we \textit{should} increase our credence in its truth.\footnote{Precisely how this should be spelled out is rather contentious, since there is a vast set of theories about the nature of dark matter which are mutually exclusive, so if we ought to believe in dark matter, which theory about it should we believe? See \cite{allzen2021scientific} for an exposition of this issue and \cite{martens2022dark} for the problems of coupling scientific realism and dark matter. See \cite{vaynberg2024realism} for a pro-realist argument for dark matter.} Such arguments can, insofar as you accept the methodological and inferential steps involved, be convincing, but the critical point is to recognize that the argument epistemically strengthens the hypothesis \textit{beyond} the accumulative or piecemeal support given by the empirical evidence itself. Invoking explanatory arguments means that you are \textit{adding} a relevant epistemic component to the total picture, but the component is \textit{not} empirical.

Explanatory arguments, or appealing to theoretical virtues, can provoke anxiety among scientists, raising concerns that theories or models such as string theory, $\Lambda$CDM, and inflation theory are based on a \textit{ philosophical} foundation rather than a scientific one. There has been, and continues to be, heated debate within the scientific community about the scientific status of such theories.\footnote{Part of the debate concerns questions which \textit{are} decidedly philosophical in nature: what demarcates science from philosophy, religion, or other human endeavors? What are the boundaries of the empirical? When is a theory true? What constitutes testability? What is confirmation? et.c. See, for example, \cite{greene2000elegant}, \cite{dawid2013string}, \cite{kragh2014testability}, \cite{ellis2014scientific,ellis2017domain}, and \cite{merritt2021cosmological,merritt2021mond}.} Although the so-called 'age of precision cosmology' marked a symbolic and significant departure from its rationalist past, elements of that past appear to have resurfaced. The issue of physicists invoking these philosophical arguments—beyond the accusations of engaging in metaphysics—is two-fold. First, constructing clear and sound explanatory arguments is a complex task, as evidenced earlier in this section. Second, the notion that explanation has \textit{anything} to do with epistemology, let alone truth, remains contentious within philosophy and carries significant theoretical baggage. Furthermore, if we accept the explanatory arguments, which aspect of the dark-matter hypothesis is epistemically enhanced? Is it the \textit{paradigm}, as Einasto suggests, or the \textit{existence} of cold dark matter, as per Hooper? Is explanatory strength probabilistic, or does it provide a reason to accept the theory as true? Why should the discovery of Neptune count as part of the inferential history of dark matter, while the failure to discover Vulcan does not? 

Although physicist engagement with the philosophical foundations of scientific epistemology is encouraging, these attempts currently lack clarity and transparency. Moreover, they appear to have emerged \emph{after} the realization that dark matter would not be detected. Like the fox in Aesop's fable, who deems the unattainable grapes sour, the sentiment seems to be that detecting dark matter is no longer necessary to confirm its existence. Alternatively, one might adopt a more optimistic metaphor: realizing that there are only lemons, one chooses to make lemonade.

\subsection{Explanation without justification}

Not all explanatory arguments carry the implication of justification. Below, Faber and Gallagher \cite{Faber_Gallagher1979}, who ‘are especially concerned with the current status of the “missing mass” problem’, uses explanatory power as a criteria for theoretical inclusion, i.e. to be included in the set of theories connected to the phenomena of flat rotation curves: 

\begin{quote}
    \small In summary, we feel that no generally valid alternative explanation has been put forward for these flat rotation curves and that the observations and their implications must therefore be taken very seriously. \cite[146]{Faber_Gallagher1979}
\end{quote}

\noindent Here, “implications” refers to the presence of additional mass (MOND had not yet been conceived at this point). That the hypothesis must “be taken very seriously” due to its explanatory advantage appears as a judgment on the \textit{viability} of the hypothesis, as opposed to its \textit{justification}. Another way to categorize this reasoning is as a no-alternatives argument, where the lack of alternative viable theories able to explain the phenomena is taken as an indication for the viability of the considered theory (in this case, the missing-mass hypothesis).\footnote{See \cite{dawid2013string,dawid2015no} for a detailed account of the no-alternatives argument.} Although the explicated examples in this paper are not sufficient to establish a strong pattern, they raise an intriguing hypothesis: explanatory arguments preceding the failure to detect dark matter are to a larger extent characterized by viewing explanatory power as an indicator of pursuitworthiness, whereas explanatory arguments succeeding this failure are characterized by considering explanatory power as truth conducive.\footnote{See \cite{SHAW2022103} for a detailed analysis of the idea of pursuitworthiness.} If true, this would entail a shift in the perception of what epistemic role and power explanations have in physics. Such a shift may seem innocent enough, but it is worth pointing out the consequences: As stated previously, believing in dark matter entails believing that it constitutes approximately 80\% of all matter in the universe. This means that the shift to using explanation as an epistemic notion carries enormous implications for what you believe the universe is fundamentally made of. 

\section{History and dark matter}

A history of dark matter already at the outset implies the existence of a thing, dark matter, about which there is a coherent history. In this sense, the mere \textit{act} of writing a history of dark matter constitutes a sort of argument for its existence. The content of that history is of great significance because, according to \cite[296]{staley2008einstein}, “scientists employ historical accounts to establish a canon and shape the boundaries of a disciplinary rereading of the past from the end of science”. In the dark matter context, the present is a natural substitute for the end, i.e. as the viewpoint from which one can interpret the past. The kind of historical account I will be concerned with may be labeled “research history” and can be defined as: 

\begin{quote}
    \small [...] explicitly historical but highly selective accounts of the emergence and implication of a theory, experiment, or discovery that are presented in major research or research review papers. \cite[297]{staley2008einstein}
\end{quote}

\noindent For example, if one's viewpoint contains a particular characterization of dark matter, a history of dark matter becomes a history of that particular characterization. Selectively presenting parts of history that support a particular characterization promotes a narrative that implies continuity in the usage and understanding of the central terms between past scientists and their present counterparts. Semantic continuity in turn implies that terms have picked out \textit{the same entity} throughout history. In the context of dark matter, this strengthens the idea that, despite substantial theoretical evolution, our knowledge about dark matter has increased. However, if there is no historical continuity, then claiming that the dark matter which \cite{Zwicky1933} inferred was confirmed by \cite{rubin1970rotation} becomes an equivocation. Worse still, if 'dark matter' as used by Zwicky referred to an entity with properties incompatible with dark matter as currently understood, then the notion of a linear history of dark matter looks more like a rereading of the past from the current viewpoint; as a path through history constructed only to meet a currently pressing epistemic need.

\subsection{The prevalence and impact of history in dark matter}

The history of a scientific theory is usually preceded by a pivotal moment in which the theory is either confirmed or falsified. The benefit of hindsight enables historians to discern the patterns which led scientists astray, when a theory has been falsified, or trace the key insights pointing towards truth, in case of its confirmation. The presence of accounts on the history of dark matter is therefore unexpected, considering that it is currently one of the biggest \textit{open} questions in physics. Even more unexpected is that a majority of these historical accounts were written by physicists actively working on the dark matter problem at the time. These histories of dark matter are not simply autobiographical memorabilia peripheral to the standing of dark matter in the broader community but are of great interest to a majority of the physicists currently working on dark matter (see fig. \ref{impact}). The high level of impact generated by a paper on the history of dark matter within the physics community tells us that it is an incredibly popular topic among its members. More importantly, however, it tells us that the content and epistemic narrative presented in a historical account may seriously influence the beliefs and credences of the community regarding the \textit{current} dark matter hypothesis. Support for the latter claim can be found by looking in cosmology textbooks, popular science media, and encyclopedic entries. It is not uncommon to come across dense historical claims in these sources, often in some version of the following proposition:

\begin{figure}[ht]
    \centering
    \includegraphics[width=0.9\textwidth]{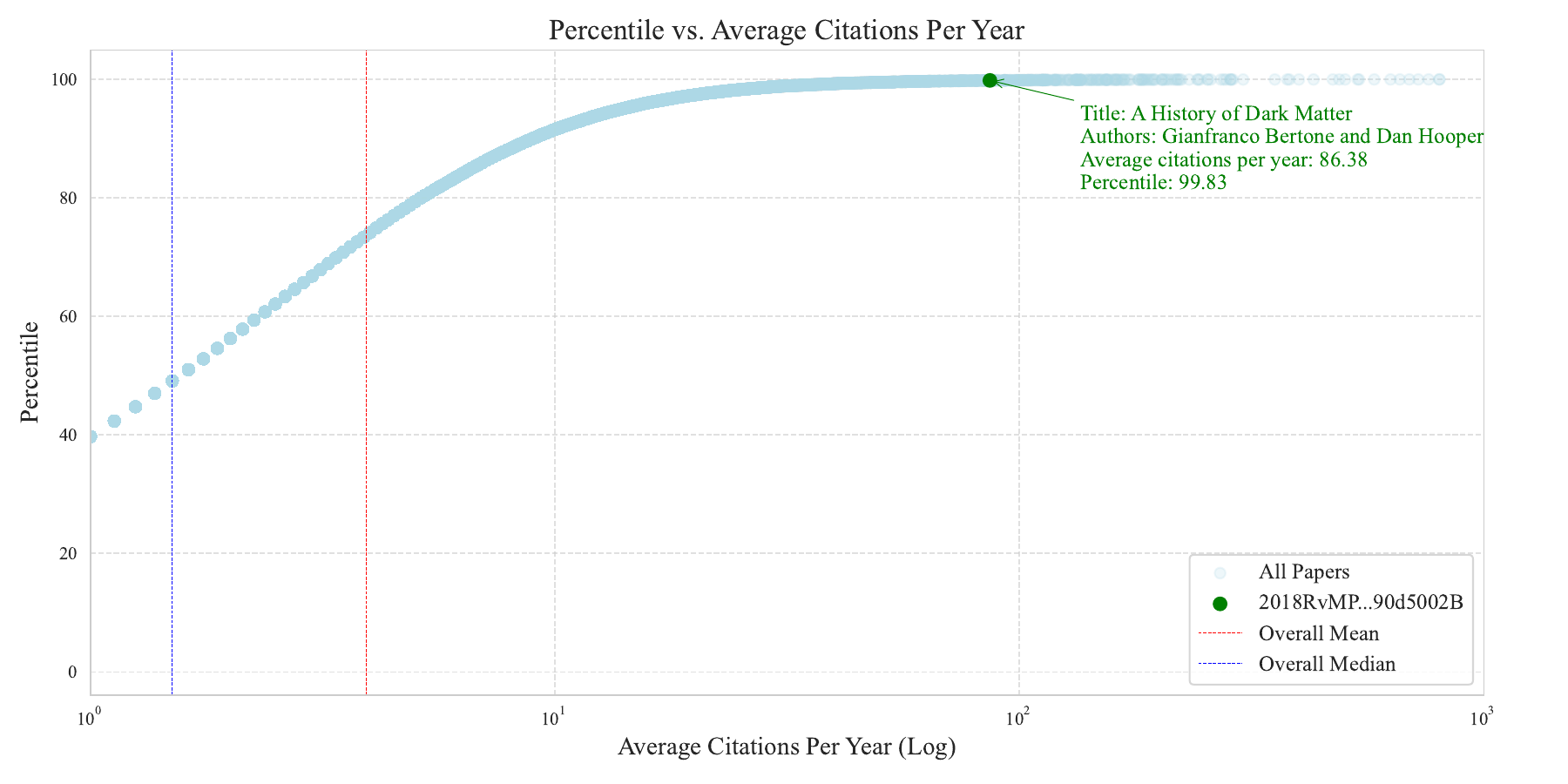}
    \caption{Plot showing the rank of papers (percentile) based on the total sum of citations a paper has divided by number of years since its publication. The highlighted paper is among the top 1\% of cited papers out of 177.284 papers mentioning 'dark matter'}
    \label{impact}
\end{figure}

\begin{quote}
    \small Originally known as the “missing mass”, dark matter’s existence was first inferred by Swiss American astronomer Fritz Zwicky, who in 1933 discovered that the mass of all the stars in the Coma cluster of galaxies provided only about 1 percent of the mass needed to keep the galaxies from escaping the cluster’s gravitational pull. The reality of this missing mass remained in question for decades, until the 1970s when American astronomers Vera Rubin and W. Kent Ford confirmed its existence [...] \cite{Encbrittanica}
\end{quote}

\noindent This nugget of dark-matter history is not to be faulted for its lack of depth, considering it is an encyclopedia entry. However, the fact that a historical description even \textit{is} included in an encyclopedia entry suggests its significance for the general perception of dark matter. The entry, though short and dense, manages to nonetheless represent the physicist perspective fairly accurately. Although the scope, depth, and details vary, the usual starting point is Fritz Zwicky's (\cite{Zwicky1933,zwicky1937masses}) studies of the Coma cluster, in which he found that luminous mass alone could not fully explain the galactic movements in the cluster. He concluded that there must be some 'dunkle materie' in the cluster, the presence of which would explain the dynamics. It is popular to credit Zwicky for introducing the term ‘dark matter’, often followed by berating the contemporary scientific community for overlooking the magnitude of Zwicky's incredible discovery. Below are three such examples:

\begin{quote}
    \small Unquestionably, Fritz Zwicky was the first to propose this form of dark matter, although for 30 years after his proposal, it was largely unappreciated; astronomers did not see Zwicky’s anomaly as a crisis leading to a possible paradigm shift. It took about 40 years for Zwicky’s insight to be fully accepted. \cite[12]{Sanders2010}
\end{quote}

\begin{quote}
   \small About 82 years ago, the Swiss astronomer Fritz Zwicky used a telescope at the Mount Wilson Observatory in California to measure the radial velocities of galaxies in the rich Coma cluster He found surprisingly large velocity dispersions, indicating that the cluster density was much higher than the one derived from luminous matter alone. He named this invisible matter dark matter, and published his findings in Helvetica Physica Acta in 1933. In the 1970s, Vera Rubin and collaborators, and Albert Bosma measured the rotation curves of spiral galaxies and also found evidence for a missing mass component. \cite[1]{baudis2016dark}
\end{quote}

\begin{quote}
    \small As Morton Roberts recounts in his short history, the evidence for something beyond stars began with Zwicky and the Coma cluster, where Zwicky described the need for dark matter to hold the Coma galaxies together. Smith showed the same was true for the Virgo cluster. Babcock measured a rising rotation curve for M31 (Andromeda) and described it in terms of a rising mass-to-light ratio. Even earlier, Lundmark found the same for M31 and a few other spirals. \cite[9]{Turner_2022}
\end{quote}

\noindent According to these historical narratives, Zwicky's work represents the genesis of dark matter.\footnote{Apart from dictionary entries and introduction chapters, this narrative is peddled in popular channels for science communication, for example the YouTube channel “The Entire History of the Universe” (episode “Where Did Dark Matter And Dark Energy Come From?” \cite{OftheUniverse2021Dec}) written with the aid of scientific advisors Dr. Joel Primack -- University of California Santa Cruz, Dr. Anatoly Klypin -- New Mexico State University, and Dr. Stefan Gottlöber -- Astrophysical Institute Potsdam.} The portrait of Zwicky as a “pioneer” {\cite[13]{bertone2018history}} or “one of those rare unorthodox geniuses” \cite[12]{Sanders2010} is a curiously frequent aspect of the narrative, the magnitude of which is sometimes striking:

\begin{quote}
    \small The discovery by Zwicky (1933) that visible matter accounts for only a tiny fraction of all of the mass in the universe may turn out to have been one of the most profound new insights produced by scientific exploration during the 20\textsuperscript{th} century. \cite[657]{van1999early}
\end{quote}

\noindent This dense and simplified history of dark matter presumably reaches the greatest number of people, influencing the perception of dark matter's history within the general population. Despite its presumed impact, the simplified version is not an interesting subject for further analysis, considering that it lacks historical accuracy by design, in order to convey the general unfolding of events. However, including it in this context is useful because it illustrates the condensed version present in what we previously labeled \textit{research history}. An example of the latter, and the subject for the remaining analysis, is a review paper written by two astrophysicists currently working on dark matter -- Dan Hooper and Gianfranco Bertone -- entitled \textit{A History of Dark Matter}. 

\subsection{A history of dynamics}

The review paper by \cite{bertone2018history} is a comprehensive and thorough account of many of the most important steps in the theoretical evolution of what is today known as dark matter. The aim is to “provide the reader with a broader historical perspective on the observational discoveries and the theoretical arguments that led the scientific community to adopt dark matter as an essential part of the standard cosmological model”. In a sense, the aim is to provide the reader with the arguments for the existence of dark matter which convinced scientists to accept it. As such, the paper should be treated as an epistemological endeavor as much as a historical one, especially considering its high impact in the dark-matter community. Their account makes extensive use of primary historical sources and state-of-the-art knowledge of dark matter to provide a comprehensive overview of dark matter's “interesting history, and how it came to be accepted as the standard explanation for a wide variety of astrophysical observations” (\cite[1]{bertone2018history}) The first part of the paper is devoted to the conceptual origins of dark matter:

\begin{quote}
    \small We study the emergence of the concept of dark matter in the late 19th century and identify a series of articles and other sources that describe the first dynamical estimates for its abundance in the known Universe. \cite[4]{bertone2018history}
\end{quote}

\noindent In the above, it is clear that the authors have already drawn a conceptual link between \textit{dynamical estimates} and dark matter. The emergence of the concept of dark matter and the dynamical estimates of it are connected in contemporary history by means of a few actors, which also serve as nodes of connection with the present:

\begin{quote}
    \small As we shall see in Chapter IV, the pioneering work of Kapteyn, Jeans, Lindblad, Öpik and Oort opened the path toward modern determinations of the local dark matter density, a subject that remains of importance today, especially for experiments that seek to detect dark matter particles through their scattering with nuclei. \cite[12]{bertone2018history}
\end{quote}

\noindent These actors are represented in the subsection “Dynamical Evidence”, which succeeds the subsection “Dark Stars, Dark Planets, Dark Clouds” in the chapter “Prehistory”. Chronologically ordered, the sections provide the reader with sources referring to a kind of proto-concept of dark matter present up until the end of the 19th century, a concept which is then narrowed by the advent of dynamical evidence in the early 20th century. Interestingly, the proto-concept includes references to black holes, included in \cite[7]{bertone2018history} “as an explicit example of a discussion of a class of invisible astrophysical objects, that populate the universe while residing beyond the reach of astronomical observations”. Interesting because \textit{invisible} and \textit{non-luminous} are not co-extensive terms, a distinction which we will see becomes significant for what parts of history one deems relevant for a history of dark matter.

In their study of the emergence of the concept of dark matter, and in opposition to the simplified history from the previous subsection, the authors emphasize that Zwicky's use of the term ‘dark matter’ was in fact neither novel nor isolated, but instead argue that the term was ubiquitous among his contemporaries:

\begin{quote}
    \small We [...] discuss the pioneering work of Zwicky within the context of the scientific developments of the early 20th century. And although his work clearly stands out in terms of methodology and significance, we find that his use of the term “dark matter” was in continuity with the contemporary scientific literature. \cite[4]{bertone2018history}
\end{quote}

\begin{quote}
    \small Although [Zwicky] doesn’t explicitly cite any article, it is obvious [...] that he was well aware of the work of Kapteyn, Oort and Jeans discussed in the previous chapter. His use of the term “dark matter” is, therefore, in continuity with the community of astronomers that had been studying the dynamics of stars in the local Milky Way. \cite[14]{bertone2018history}
\end{quote}

\noindent This account rejects the narrative trope of Zwicky as a lone and misunderstood genius, and with it the notion that he was the first to “discover” that additional matter was needed to explain the observed phenomena. Instead, the authors stress that Zwicky's inference to low-luminous matter, as well as his use of the phrase 'dark matter', conformed to the inferential and terminological practices established in the contemporary scientific community. This community is then said to be constituted by Kapteyn, Oort, and Jeans, i.e. the actors present in the section on dynamical evidence. Zwicky's results are taken to align with \cite{SMITH1936CMWCI.532....1S}, who published his estimates of the mass of the Virgo Cluster based on its dynamics. These estimates were later cited by Hubble in \textit{The Realm of Nebulae}. The account says little of the development in the years running up to the late 1960's, suggesting little or no significant work on the topic until then.
This narrative draws two implicit lines through history. The first line is drawn between contemporary instances of the phrase 'dark matter', which serves to establish that its historical use was coupled to or characterized by \textit{ explanations of dynamical phenomena}. This particular characterization is naturally substantiated by reference to the ubiquity of its usage in the contemporary astronomical community. The second line traces the conceptual legacy of this characterization of dark matter throughout the 20th century, all the way to our modern characterization of dark matter, implying that the latter is directly descending from the former. The conceptual continuity of dark matter has been preserved by the persistence of its semantic nucleus. 

If one can establish the historical presence of semantic continuity for 'dark matter', this fact can also serve as justification for the continuity of its \textit{ontology}. If past astronomers by 'dark matter' meant something like “invisible mass influencing the dynamics of visible systems”, then the \textit{meaning} of dark matter has remained sufficiently intact throughout history. And if the meaning of the phrase 'dark matter' has remained intact, it can be argued that scientists throughout history must have referred to the \textit{same thing} by using it. It becomes suggestive to think of the dark-matter hypothesis as self-identical over time, despite the extent of its theoretical and conceptual evolution.

In what follows, I will argue that the purported semantic continuity of the term 'dark matter' is an artificial construct created to highlight a path through history only visible through the lens of a modern definition of dark matter.

\subsection{A different history of dark matter}
 Although I agree with \cite{bertone2018history} that Zwicky's use of 'dark matter' should be seen as in continuity with the established terminology in contemporary scientific literature, I disagree with their characterization of the content of that terminology. The authors' description of how the concept of dark matter was understood by Zwicky and his contemporaries is an artifact originating from present ideas about dark matter. Kapteyn, Oort, and Jeans are taken as representative of Zwicky's contemporary community only because the content of their work aligns with the idea that dark matter is inexplicably connected to \textit{dynamical phenomena} -- a connection which is not only compatible with, but incidentally also central to, the modern characterization of dark matter. The presence of semantic continuity in early 20th century astronomy can be determined by examining the contemporary literature. Should uses of the term 'dark matter' be found in disagreement with the characterization given by \cite{bertone2018history}, the idea of semantic continuity \textit{as presented}, becomes less plausible and its presence forced.

\begin{figure}[ht]
    \centering
    \includegraphics[width=0.58\textwidth]{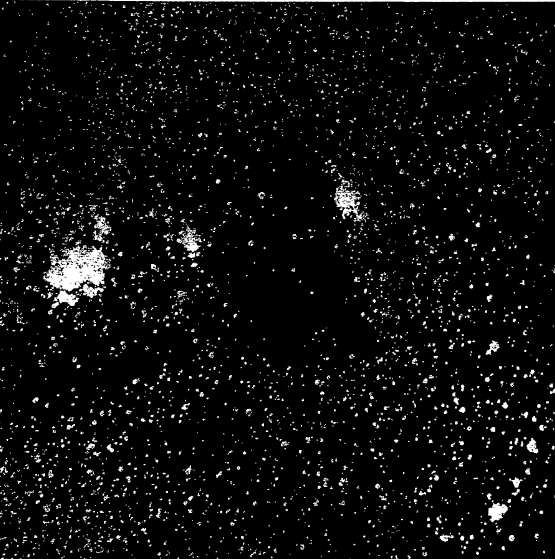}
    \caption{Dark nebula Barnard 86, depicting a “hole” in the Sagittarius star cloud. \cite{Trumpler1932}}
    \label{darkhole}
\end{figure}

An example which speaks against semantic continuity comes from another Swiss-American astronomer, Robert J. Trumpler. In the year before the publication of \cite{Zwicky1933}, \cite{Trumpler1932} published the paper \textit{Dark Nebulae} that attempts to explain an astronomical phenomenon; a dark hole which had been captured in a photograph of the Sagittarius cloud of the Milky Way (see fig.\ref{darkhole}).\footnote{Trumpler builds on earlier work from \cite{Barnard1919}, who states that he 'do not think it necessary to urge the fact that there are obscuring masses of matter in space. This has been quite definitely proved in my former papers on this subject. If any doubt remains of this it will perhaps be readily dispelled by a close examination of the photographs previously printed”.} Throughout the article, Trumpler refers to the cause of this phenomenon as 'dark, opaque material'; 'obscuring masses'; 'dark obstacles'; 'dark matter'; and 'dark stuff'. He suggests that the cause of the phenomena could be the presence of matter which \textit{interfered} with the light coming from the otherwise dense region of stars, but was not itself luminous, causing it to appear as a dark spot or hole. Here, Trumpler speculates that this 'dark matter' could constitute a novel state of matter:

\begin{quote}
    \small The matter constituting our universe is evidently found in either of two states: In organized bodies like the Sun and the stars, which by their peculiar regular and symmetrical constitution have reached a stage of luminous radiation and cheer our eyes with their twinkling light; or in unorganized, chaotic masses of tiny particles irregularly scattered through vast space, mostly dark, only in few places becoming visible as nebulae. \cite[182]{Trumpler1932}
\end{quote}

\noindent Trumpler's dark matter is characterized by its ability to interact with light, because this is the property necessary to explain the phenomena. We see here how the distinction between \textit{invisible} and \textit{non-luminous} may impact the selection of the history. The most important point to stress here is that Trumpler's dark matter is \textit{contemporarily compatible} with the dark matter of Oort, Kapteyn, Jeans, and Zwicky. There were no reasons at the time to distinguish dark matter as described by Trumpler from dark matter described by, for instance, Oort or Zwicky: that dark matter could interfere with light would have been of no significance with respect to its ability to explain dynamics. The reason for excluding Trumpler as a representative of the contemporary astronomical community can be argued to be based on the fact that his characterization of dark matter is incompatible with the current dark-matter concept.
Trumpler's conception of dark matter is very briefly presented in \cite[9]{bertone2018history} as part of the \textit{proto-concept} of dark matter, a concept which belonged to the 19th century but did not extend into the 20th century. This narrative could have been plausibly argued for had Trumpler's usage been an outlier. However, examples of the use of 'dark matter' as an explanation for light-obscuring phenomena are extensive and importantly not distinguished from the use of 'dark matter' as an explanation of dynamics.

\subsection{The dual definition of dark matter in early 20th century astronomy}

\noindent The historical focus on dark matter as an important component in dynamics overlooks its use in explanations of light-obscuring phenomena. The hypothesis that the areas of the night sky that appeared devoid of stars could be caused by matter which obscured light had, according to \cite{1904PA.....12..509W}, been around already since 1891. The hypothesis was:

\begin{quote}
    \small strongly advocated some thirteen years ago by Mr. A. C. Raynard, then Editor of \textit{Knowledge}, but is not generally accepted among astronomers.
\cite[215]{1904PA.....12..509W}
\end{quote}

\noindent In \cite{Lundmark1921PASP...33..324L}, the Swedish astronomer Knut Lundmark refers to the hypothesis of light absorption as a possible explanation of the observed 'dark lanes' found in photographs of the spiral galaxy (then nubulae) known as Messier 33. He conjectures that the appearance of these dark lanes could result from:

\begin{quote}
    \small whorls of dark matter between whorls of nebular matter and much pictorial evidence is in favor of this idea. \cite[324]{Lundmark1921PASP...33..324L}
\end{quote}

\noindent In a later paper, Lundmark had become a little less speculative about the existence of this dark matter, as evidenced when he, after calculating the total mass of the visible stellar universe, added this disclaimer to his estimate: “Of course, dark stars and dark matter exist and increase that value” \cite[896]{Lundmark1925MNRAS..85..865L}. 5 years later, \cite{Lundmark1930} include dark matter in estimations of the total mass in galactic systems, an estimate determined by the ratio between luminous mass and dark matter, a ratio which he derived using spectrographic data of the rotational velocities of galaxies (see \ref{lundmark}). In Lundmark's body of work, we see how dark matter is used both as an explanation for light absorption \textit{and} galaxy dynamics. \cite{bertone2018history} shares a glimpse of this in the section on galactic rotation curves, writing that “Holmberg argued in 1937 that the large spread in mass-to-light ratios found by Lundmark was a consequence of the absorption of light “\textit{produced by the dark matter}” \cite[18-9]{bertone2018history}. The “duality” of dark matter usage is not only present in Lundmark but also finds support in other sources. In 1927 Dutch astronomer Anton Pannekoek explicitly used the dual function of the term when attempting to account for the mass needed in the galactic nucleus, should the hypothesis that our galaxy is a rotating spiral system be true:

\begin{figure}[ht]
    \centering
    \includegraphics[width=0.7\textwidth]{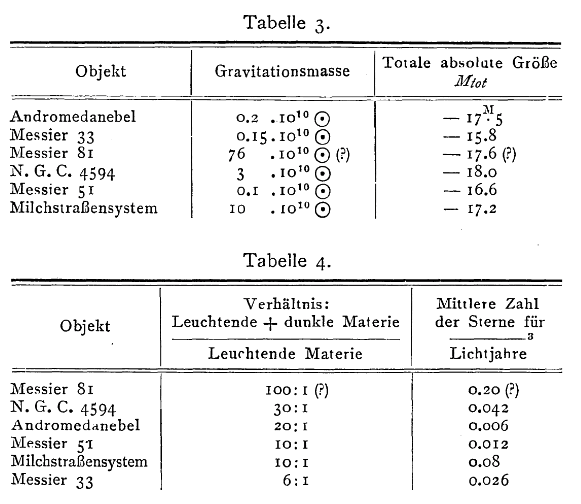}
    \caption{From \cite{Lundmark1930}. Table 4 describes the relative presence of luminous vs. dark matter in six galactic systems, including the Milky Way.}
    \label{lundmark}
\end{figure}

\begin{quote}
    \small Thus we come to the conclusion that visible stars of the galactic system cannot provide the required central attracting masses. [...] Cannot the dark matter itself act as a part of the attracting masses? In the visual aspect of the Milky Way the dark absorbing nebulae have the same importance as the luminous star clouds. \cite[40]{1927Pannekoek}
\end{quote}

\noindent Two years later, Pannekoek's idea is picked up by \cite{Harper1929JRASC..23..119H} in his president's annual address for the Royal Astronomical Society of Canada. Included in his summary of the “current progress in astronomy” is a discussion on the “Rotation of Galaxy”. The discussion centers around the work of Strömberg and Lindblad, who had hypothesized that our galaxy was a rotating spiral nebula, which would explain the observed asymmetric velocities of stars. Harper refers to a pair of papers published by Oort which are “tending to verify the soundness” of the rotation hypothesis.\footnote{From Oort, \cite[270]{Oort1928BAN.....4..269O}: “Of the various objections against the tentative [rotational] model of the galactic system the most serious one is perhaps that a very excentric position of the sun, as implied by the theory, is contradictory to observations of the density distribution of the stars. It certainly would be so if we assumed that the light of all the stars reaches us without obstruction. We know, however, that even in our immediate neighborhood considerable obstruction does take place in extended regions covered by dark nebulae. The above difficulty may possibly be only apparent and caused by the observed density distribution in the direction of the galaxy being as much determined by the distribution of of dark matter as by that of the stars themselves.”} One implication of the hypothesis is that the mass “required at the centre to give the requisite rotational velocity” is estimated at around $8\times10^{10}$M$_{\odot}$, prompting Harper to question the evidence for the presence of such a mass:  

\begin{quote}
\small This central nucleus is situated in the plane of the galaxy and in galactic longitude approximately $325^\circ$ in the general direction of the constellation \textit{Sagittarius}. Here the Milky Way clouds are brighter than in other parts of the sky, and Pannekoek examines their total luminousity to see if they would yield the necessary mass. He is forced to admit that [...] the visible stars of the galactic system cannot provide the central mass required. The question arises, would not the dark matter in our system yield this attracting mass? Photographs of distant spirals seen edge-on suggest enormous masses of obscuring material. Similar evidence is afforded nearer home in the almost complete obliteration of faint stars in certain regions of the sky by occulting matter which must be extremely tenuous but nevertheless of great effect when summed up over large volumes of space. \cite[124-5]{Harper1929JRASC..23..119H}  
\end{quote}

\noindent The idea of turning to dark matter when one has a deficit in the mass-density budget is evidently not new. The contemporary literature with Zwicky contains an abundance of examples that illustrate the ubiquity of using the term 'dark matter' in explanations of phenomena related to both light absorption and dynamics. As Harper notices, even in Oort this understanding seems dominant. For Oort, estimating the mass, velocity, and rotation of a stellar system depended on assessing the luminosity of that stellar system reliably, but the existence of light absorption due to dark matter introduced unreliability in those assessments. In a public lecture entitled \textit{Non-Light Emitting Matter in the Stellar System}, given at Leiden on the occasion of his appointment as \textit{privaatdocent}, \cite{oort1927nietlichtgevende} cites Lundmark and uses dynamical calculations to determine the amount of dark matter, which in turn is used to determine the degree of light absorption.\footnote{Roughly, a privaatdocent was at the time a teaching position connected to new academic fields.} 

\begin{quote}
    \small Do we then have to give up hope to learn about the absorption in the interstellar gas? Not entirely. [...] I have told you above that \textit{the observed velocities of the stars allow us to make an estimate of the mass of the non-light-emitting matter}. [...] It is not unreasonable to assume that apart from the light-emitting stars there could be a large quantity of dark bodies in the universe. If these bodies would be in the form of very faint stars of the same size and mass as the observed stars, then it is easy to show that there cannot be enough of these dark stars to provide an observable extinction. It becomes completely different, when the available mass would be in the form of small solid particles (such as grains of sand, for example), distributed throughout the Galactic System. We then have just about enough to give rise to a significant absorption over distances of a thousand light-years and we can easily assume these particles to be spread out so densely over a large area of space that more distant stars are completely extinguished. \cite[61-2]{oort1927nietlichtgevende}
\end{quote}

\noindent Oort is clearly using his dynamic calculations of dark matter to estimate the amount of light absorption that one can expect from it. The two phenomena must be manifestations of the same cause for dynamic calculations to be informative about the amount of light-absorption. The Estonian astronomer Ernst \cite{Opik1929PKUJ...27F...1.} concludes that “in some cases absorption by dark matter appears probable” as an explanation of obscured regions in space. The idea that light-absorption obstructed the estimations of mass from luminosity lead Trumpler (\cite[1]{TRUMPLER1930PASP...42..214T}) to suggest that the “absorption of light is thus intimately related to the question of the presence, distribution, and constitution of dark matter in the universe”. Dutch astronomer and mathematician Willem de Sitter (\cite[91]{deSitter}) writes that “evidence has accumulated” in support of the idea that “clouds of dark matter [...] prevent us from seeing the stars”.\footnote{See also: \cite[49]{NotesBritish1930JBAA...41...45.} and \cite[342]{Markvov}}

The above examples suffice to establish that the contemporary community with which Zwicky was in continuity had a much wider understanding of the concept of dark matter than alluded to in previous historical accounts. This wider concept remained established in the community at least until the late 1950s, as illustrated by its presence in \cite[48]{1939Wallenquist}, \cite[161--2]{1941Trumpler}, \cite[8]{Lindblad1948StoAn..15....4L}, \cite[20]{Lindblad1949StoAn..15....9L}, \cite[8]{1950Lindsay}, \cite[60]{1951Lindblad}, \cite[235]{1952Oort}, and \cite[8]{1952Holmberg}. The characterization and longevity of this wider concept has two consequences. First, it undermines the notion that the core characteristic of the semantic continuity of the term 'dark matter' in the first half of the 20th century can be understood only in the context of explanations for dynamical phenomena. Second, it raises doubts about the prospect of construing the hypothesis of dark matter as a single coherent historical theoretical framework. If we should consider the historical concept of dark matter as part of the history of dark matter, we have to accept that it is semantically discontinuous and in ontological contradiction with the modern concept of dark matter.

Despite the increased historical resolution provided by \cite{bertone2018history}, the narrative image remained the same: the history of dark matter is presented as coherent and continuous, where evidence steadily accumulates toward the modern dark-matter hypothesis. However, the historical dots between which the line through history is drawn can be identified by their consistency with the current understanding of dark matter.

\section{Concluding remarks}
The lack of conclusive confirmation of the existence and nature of dark matter in the form of detection has led some to seek epistemic justification elsewhere. Explanatory arguments, theoretical virtues, and historical narratives have all been recruited in attempts to rationalize the acceptance of the reality of dark matter in the scientific community. For philosophers of science and physics, the two former are part of their core area of expertise, and for historians of science and physics the same is true of the latter. Despite the available expertise, physicists have, to a large extent, taken it upon themselves to be philosophers and historians, with mixed results to show for it. 

With respect to explanation, it is not sufficient to argue that dark matter displays explanatory power, or that its introduction unifies a variety of phenomena. One has to engage in the wider epistemology of explanation for such arguments to be significant. What are the reasons for believing that explanatory power is a guide to truth? Is the ability to explain connected to empirical success? Is explanation part of the justificatory framework, or should it be recognized as a reason to pursue a theory? Explanation is a potentially powerful epistemic notion, but should be approached with greater care and used more thoroughly to be harnessed in relation to dark matter. In this regard, increased interdisciplinary work in cooperation with philosophers of science and physics could generate more robust arguments.

With respect to the history of dark matter, it appears closely related to the phenomena it has been invoked to explain. In a sense, the term dark matter is dynamic because its definition is fixed by and relative to the contemporary phenomena it is assumed to cause. It is in response to the progression of science that the evolution of the use of 'dark matter' can be traced. Our current unexplained phenomena dictate the specification of dark matter as a non-baryonic particle that is electromagnetically inert and exists in specific quantities and distributions. Because the unexplained phenomena of the past were different, the specifications of dark matter were different. In addition to influencing the dynamics of galactic systems, it also had to absorb or interfere with light. The advancement of observational technology revealed that dense cosmic dust caused light absorption, thereby removing the need to explain it with dark matter. Since the property of light interaction was no longer required to explain the remaining unexplained phenomena, it was omitted from the meaning of dark matter. This explains why, when writing the history of dark matter, the tendency is to only see parts of the history that reflect the current meaning.

\begin{quote}
\small Like intellectuals and academics in all fields, at the beginning of their books or articles physicists tend to create a frame in which to insert their own research program. They tend to build a narrative; to draw a coherent itinerary through the past, an itinerary which links their own questions and solutions to the ones previously debated, and currently accepted, by at least part of the community. \cite[203]{pestre1999commemorative}
\end{quote}

\begin{quote}
    \small If modern science functions as a mark book for earlier science, one will tend to present occurrences that can be seen today as pioneering as though they were just as pioneering in their historical situation. And one will evaluate the knowledge of the past as though it concerned the same subject and concepts that we think it was 'really' about today. \cite[94]{kragh_1987}
\end{quote}

\noindent Current efforts to understand the history of dark matter are underdeveloped, but its increasing popularity among academics in physics, philosophy, and history shows great promise in shedding light on questions of theory evolution in science.

\section*{Acknowledgments}
\small I'm grateful to Dan Söderberg, who translated Knut Lundmark's 1930 paper from its original German to English; to audiences in Amsterdam, Utrecht, and Bristol; and to Jeroen van Dongen, Robert van Leeuwen, Richard Dawid and Karim Thebault for helpful comments. Funding for this research was provided by the National Swedish Research Council (Vetenskapsrådet) grant number 2022-06143.

\bibliography{Ref}
\bibliographystyle{unsrt}

\end{document}